\documentclass[a4paper,11pt]{article}
\pdfoutput=1 

\usepackage{jinstpub} 
\usepackage{siunitx}
\usepackage{wasysym}
\usepackage{xspace}
\usepackage{subfigure}

\title{\boldmath Ultra-low material pixel layers for the Mu3e experiment}

\author[a]{N.~Berger,}
\author[b,1]{S.~Dittmeier,\note{Corresponding author.}}
\author[b]{L.~Henkelmann,}
\author[b]{A.~Herkert,}
\author[b]{F.~Meier~Aeschbacher,}
\author[b]{Y.W.~Ng,}
\author[b]{L.O.S.~Noehte,}
\author[b]{A.~Sch\"oning}
\author[b]{and D.~Wiedner}

\affiliation[a]{Institut f\"ur Kernphysik, Johannes Gutenberg-Universit\"at Mainz,\\Johann-Joachim-Becher-Weg 45, 55128 Mainz, Germany}
\affiliation[b]{Physikalisches Institut, Ruprecht-Karls-Universit\"at Heidelberg,\\Im Neuenheimer Feld 226, 69120 Heidelberg, Germany}

\emailAdd{dittmeier@physi.uni-heidelberg.de}

\abstract{
The upcoming Mu3e experiment will search for the charged lepton flavour violating decay of a muon at rest into three electrons. 
The maximal energy of the electrons is \SI{53}{MeV}, hence a low material budget is a key performance requirement for the tracking detector. 
In this paper we summarize our approach to meet the requirement of about \SI{1}{\permil} of a radiation length per pixel detector layer. 
This includes the choice of thinned active monolithic pixel sensors in HV-CMOS technology, ultra-thin flexible printed circuits, and helium gas cooling.
}


\collaboration[c]{on behalf of the Mu3e collaboration}

\proceeding{8$^{\text{th}}$ International Workshop on Semiconductor Pixel Detectors for Particles and Imaging\\
  September 5 - 9, 2016\\
  Sestri Levante, Italy}

\newcommand{\mteee}{$\mu^+ \rightarrow e^+e^-e^+$\xspace}
\newcommand{\mtenunu}{$\mu^+ \rightarrow e^+e^-e^+\nu_e\bar{\nu}_\mu$\xspace}

\newcommand{\kapton}{Kapton\textsuperscript{\textregistered}\,}

\begin{document}
\maketitle
\flushbottom

\section{The Mu3e experiment}
\label{sec:intro}

\begin{figure}[tb]
	\centering
	\includegraphics[width = .86\textwidth]{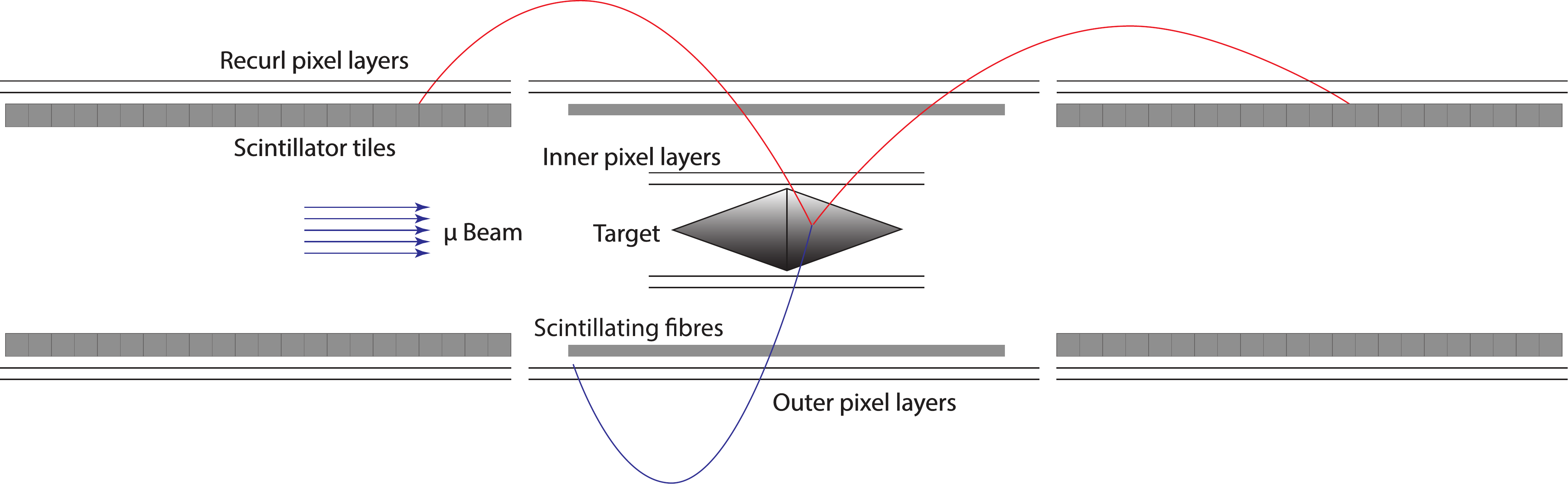}
	\caption{Schematic of the Mu3e detector.}
	\label{fig:mu3e_detector}
\end{figure}

The Mu3e experiment~\cite{Mu3e} aims to search for the charged lepton flavor violating decay \mteee with a sensitivity to a branching ratio of \num{e-16} (\SI{95}{\%} C.L.) $-$ 
four orders of magnitude smaller than the current branching ratio limit of \num{e-12} set by the SINDRUM experiment~\cite{SINDRUM}. 
To achieve this goal within a reasonable time, more than \num{e9} muon decays per second have to be observed. 
An excellent momentum resolution of $\sigma_p \approx \SI{0.5}{MeV/c}$ is required to distinguish between the signal decay and the dominant source of background, the decay \mtenunu~\cite{Djik}. 
Tracking detectors based on silicon pixel sensors are most suitable to meet the requirements on rate and resolution. 
As multiple scattering limits the momentum resolution, the whole detector has to be very thin. 
The material budget is limited to \SI{1}{\permil} of a radiation length ($X_0$) per layer. 
Hybrid silicon pixel detectors offer high rate capabilities, however, the material budget is intrinsically too high as sensor and readout ASIC are two independent components. 
The high voltage monolithic active pixel sensor~\cite{Peric:2007zz} (HV-MAPS) combines high speed and low material budget ($\SI{0.5}{\permil}\,X_0$ if thinned down to \SI{50}{\micro\metre}) and is therefore chosen as technology for the experiment. 
A schematic of the planned barrel shaped tracking detector together with the scintilliating fibre and tile detectors used for precise time measurements is shown in figure~\ref{fig:mu3e_detector}. 
In this paper, the timing detectors will not be further discussed. 

\begin{figure}[b]
	\centering
	\subfigure[][ 	\label{fig:module}]{\includegraphics[width = .45\textwidth]{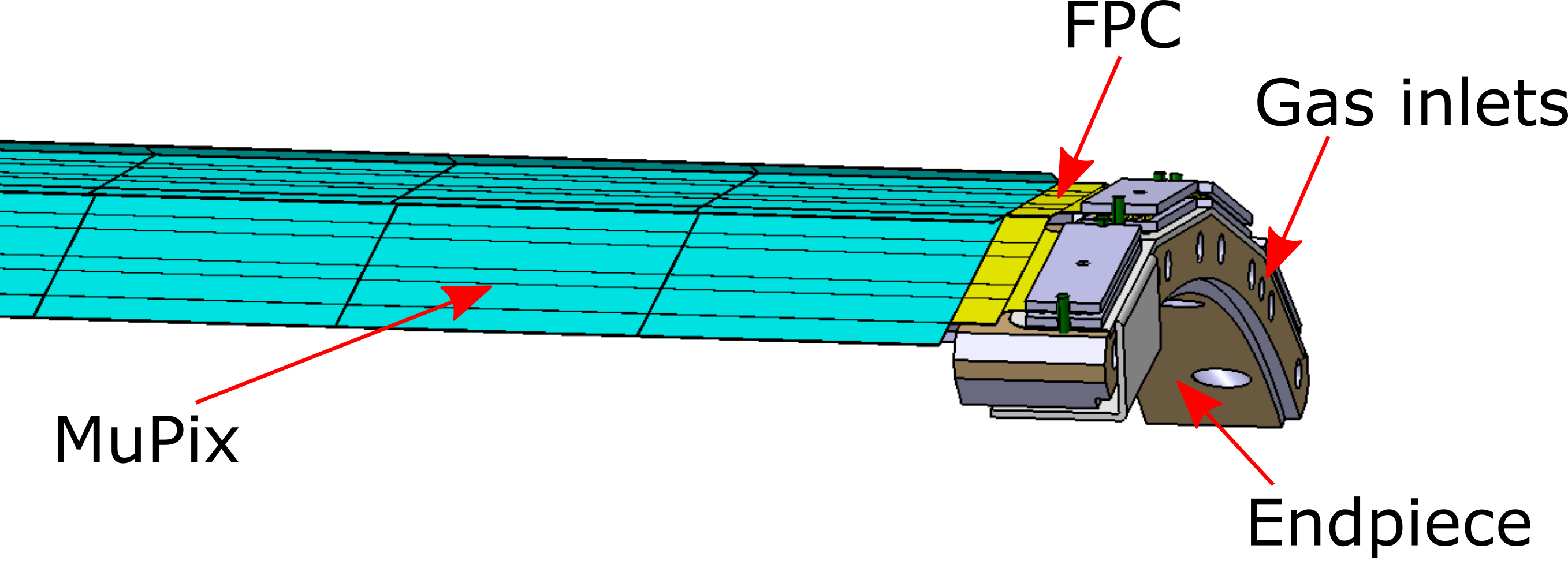}}
	\qquad
	\subfigure[][	\label{fig:module_bottom}]{\includegraphics[width = .45\textwidth]{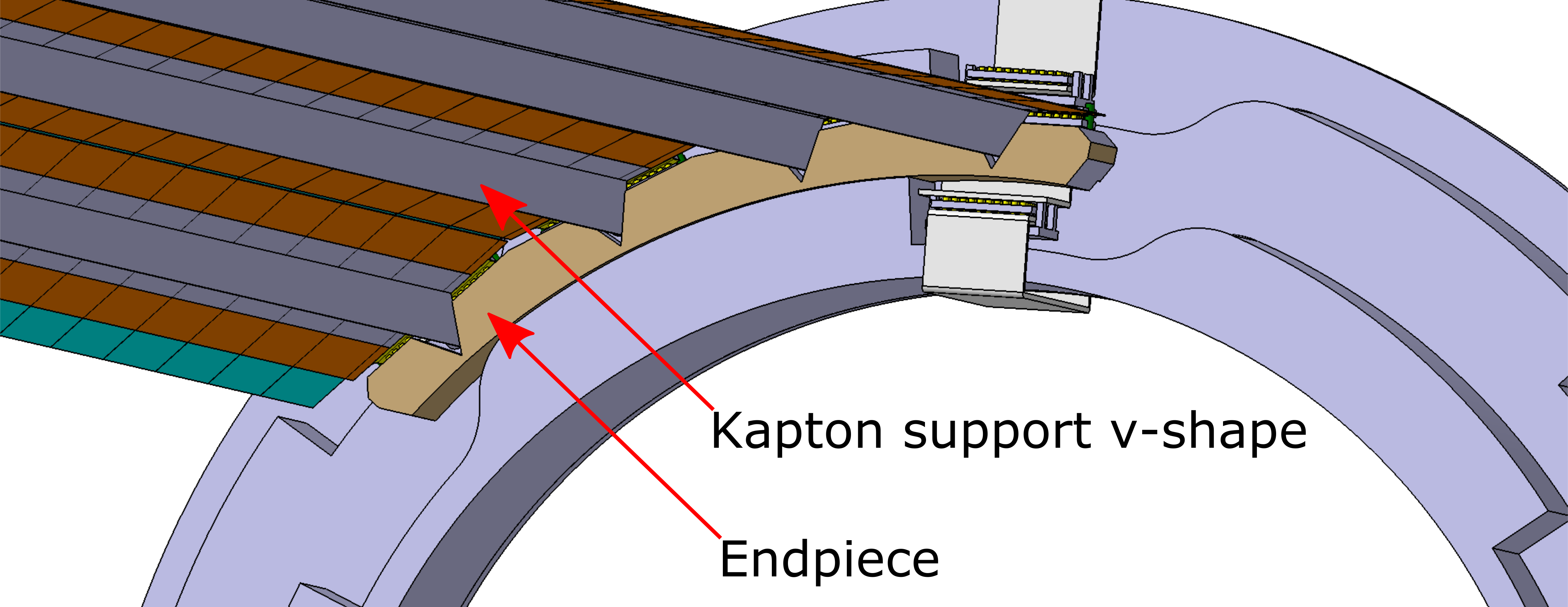}}
	\caption{Pixel detector modules for Mu3e. (a) Rendering of a pixel module for the Mu3e detector. 
	The sensors (MuPix) are connected to the FPC which is mounted on the endpiece. 
	The endpiece has gas inlets for the helium cooling. 
	(b) Bottom view of a mounted module. 
	The v-shaped \kapton support allows for an additional helium gas flow close to the sensors.}
	\label{fig:modules}
\end{figure}

To meet the strict material budget requirement, the material of the interconnects, the mechanics and the cooling system inside the active detector volume has to be minimized. 
Therefore, thin aluminium flexible printed circuits (FPC) ($\approx \SI{0.5}{\permil}\,X_0$), a mechanical support made from \kapton foil ($\approx \SI{0.1}{\permil}\,X_0$), and gaseous helium cooling are foreseen. 
The estimated peak material budget is about $\SI{1.15}{\permil}\,X_0$ per tracking layer, taking up to \SI{20}{\micro\metre} of glue into account. 
A drawing of a Mu3e pixel module is shown in figure~\ref{fig:module}. 
A bottom view, where the v-shaped support structures can be seen, is depicted in figure~\ref{fig:module_bottom}. 
These v-shapes increase the mechanical stability and allow for additional helium gas flow close to the sensors. 

The FPC technology currently under consideration and feasibility studies of an FPC prototype are presented in section~\ref{sec:fpc}. 
In section~\ref{sec:he}, the cooling concept for the Mu3e pixel detector is discussed. 

%


\section{Flexible printed circuits}
\label{sec:fpc}

The FPC provides all necessary connections for the pixel sensors.
This includes the supplies (power, ground, and bias voltage) and signals (clock, reset, slow control and up to three fast data transmission lines). 
For the inner layers (\num{6} sensors, \SI{12}{cm} long), each sensor has three LVDS links at an output data rate of \SI{1.25}{Gb/s}.
For the outer layers (\num{18} sensors, \SI{36}{cm} long), only one data link per sensor is required due to the lower hit occupancy. 
The detector layers are electrically split in half with connectors at both ends. 
In order to minimize the material budget, aluminium is the preferred conductor material, saving a factor of four in material compared to copper for the same conductivity. 




\subsection{Technology}

We currently explore two options for the FPCs. 
The first technology, called HiCoFlex~\cite{HighTec} (by HighTec, Switzerland) uses very thin copper layers (few \si{\micro\metre}) and allows for structure sizes down to a few \si{\micro\metre}. 
Our second option is an FPC manufactured by LED Technologies of Ukraine (LTU)~\cite{LTU}. 
Foils with \SI{14}{\micro\metre} of aluminium on \SI{10}{\micro\metre} of \kapton are the raw material for the electrical layers. 
The FPC has two electrical layers separated by a dielectric spacing of \SI{45}{\micro\metre} made of \kapton and glue required to match the differential impedance to $Z_{diff} = \SI{100}{\ohm}$. 
The minimum structure size is \SI{65}{\micro\metre}. 
The total FPC has an estimated thickness of \SI{83}{\micro\metre}, corresponding to a peak material budget of about $\SI{0.5}{\permil}\,X_0$. 
Bonding to sensors or PCBs as well as layer-to-layer vias are realized using the Single point Tape Automated Bonding (SpTAB) technique~\cite{SpTAB}. 
The advantage of this bonding method is that neither additional high Z materials nor fragile bonding wires are necessary. 

%

\subsection{Feasibility studies}

\begin{figure}[b]
	\centering
	\includegraphics[width = .7\textwidth]{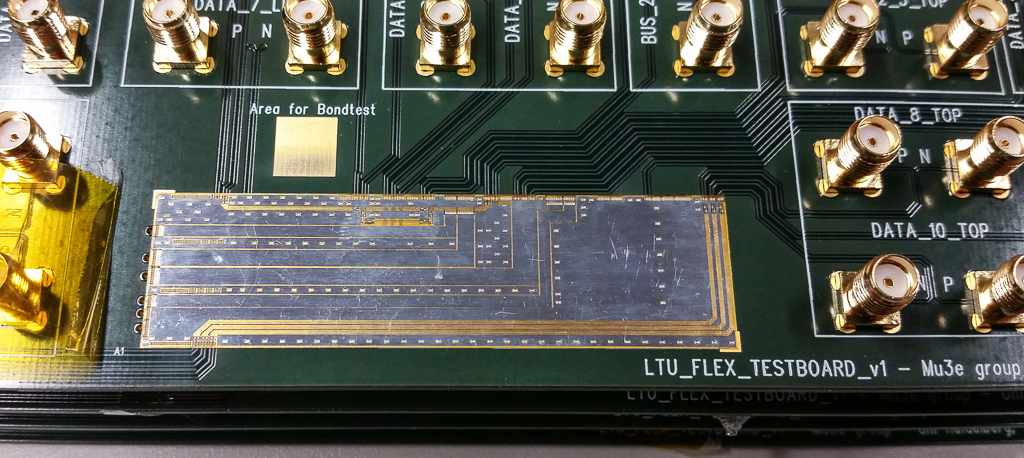}
	\caption{The FPC prototype with test structures bonded to a PCB. The FPC is \SI{6.85}{cm} long and \SI{1.9}{cm} wide.}
	\label{fig:testflex}
\end{figure}

We performed feasibility studies with an FPC by LTU~\cite{bsc:noehte}. 
To this end we designed an FPC with test structures which was produced and bonded to a test PCB, see figure~\ref{fig:testflex}. 
Five identical prototypes were tested. 
The actual thickness of the FPCs was measured with a thickness gauge to be \SI{85}{\micro\metre} on average over the whole FPC, ranging from \SIrange{77}{95}{\micro\metre}. 
The individual glue layers are estimated to have a thickness of \SI[separate-uncertainty]{6(3)}{\micro\metre}, approaching the desired \SI{5}{\micro\metre} quite well. 
For a better control of the variations, the development of a special gluing jig is foreseen. 
  
The resistance of dedicated low resistance power lines was measured using four-wire sensing. 
The results, between \num{50} and \SI{130}{\milli\ohm} depending on the trace length, are compatible with a conductor thickness of \SI[separate-uncertainty]{12.2(3)}{\micro\metre}, significantly lower than \SI{14}{\micro\metre}. 
As the oxide layers of aluminium are typically only a few ten nanometers thin~\cite{Jeurgens200289}, they do not account for this reduction. 
Hence, we assume that aluminium is most likely lost during production in the etching process. 

We tested the quality of high speed differential data transmission lines by using a bit error rate test. 
A PRBS7 pattern was generated and checked using an Altera Stratix V FPGA. 
At the experiment's target data rate of \SI{1.25}{Gb/s} we did not observe any errors in ten tested data traces over several hours, resulting in an upper limit on the bit error rate at the \SI{e-13} level. 
Up to \SI{2.5}{Gb/s} no bit errors were observed on a similar timescale, showing that the bandwidth margin is large. 
%

The impedance of the differential transmission lines was analyzed using a time domain reflectometer. 
The traces were designed to match a differential impedance of \SI{100}{\ohm} using a trace width of \SI{63}{\micro\metre} and a gap of \SI{133}{\micro\metre}. 
We found that the impedance of most of the data traces is off by more than \SI{10}{\%} from the target. 
On the bottom layer, 
the distance between FPC and the first copper layer on the board was not taken properly into account in the design resulting in an impedance between \SIrange{85}{92}{\ohm}. 
On the top layer, 
the impedance is measured to be higher than \SI{100}{\ohm} (around \SIrange{120}{130}{\ohm}) as the \kapton foil, foreseen to be used for mechanical support in the experiment, was included in the design but not glued to the FPC for the measurements. 
Between the five different FPCs no significant variations are observed.
The observed deviations are within expectations and are due to limitations of the experimental setup. 
For the final FPC the parameters will be adjusted to match the requirements. 



\section{Helium gas cooling for Mu3e}
\label{sec:he}

\begin{figure}[tb]
	\centering
	\includegraphics[width = .85\textwidth]{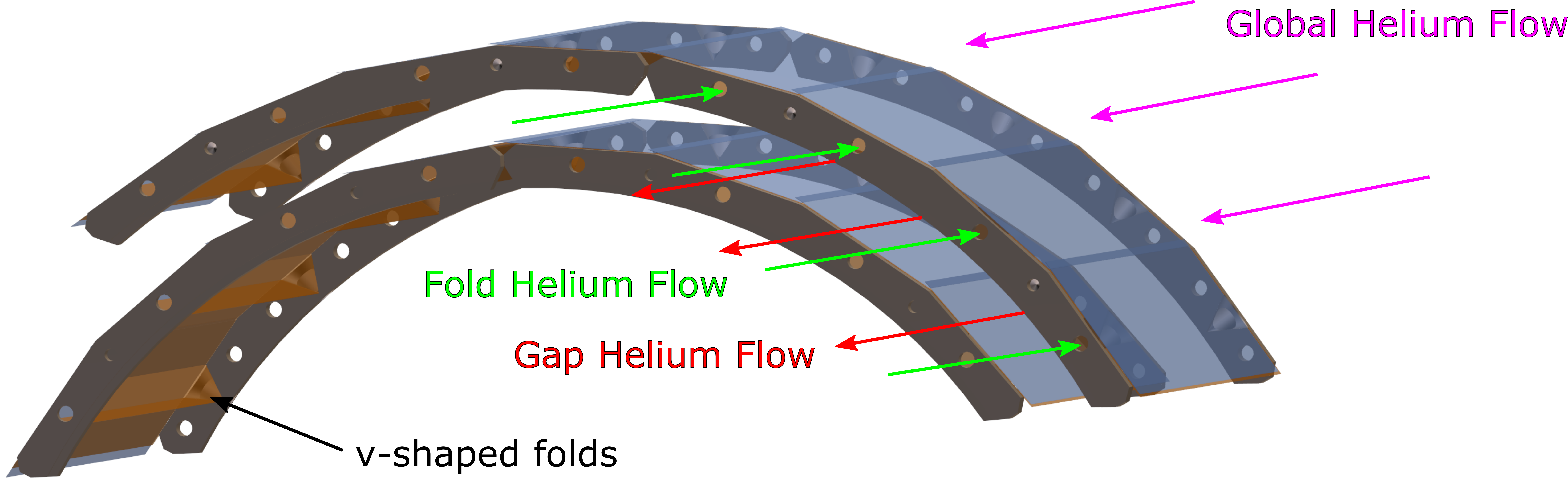}
	\caption{The helium gas cooling concept for the Mu3e pixel detector, visualized using a CAD model of the outer detector layers. 
	The whole pixel detector is located inside a flowing helium atmosphere (global flow, purple arrows). 
	In the gaps between layer \num{1} and \num{2}, and between layer \num{3} and \num{4}, a helium flow (gap flow, red arrows) is introduced by gas inlets in the barrels' endrings. 
	In the support structure of the outer layers v-shaped folds, see also figure~\ref{fig:module_bottom}, allow for an additional flow in the opposing direction (fold flow, green arrows).}
	\label{fig:cooling_schematic}
\end{figure}

With a maximum power consumption of \SI{400}{mW/cm^2} by the pixel sensors, the Mu3e silicon pixel detector will produce up to \SI{4.7}{kW} of heat, which has to be dissipated. 
The task of the Mu3e cooling system is to cool the sensors below \SI{70}{\celsius} and keep the material budget as low as possible. 
As gaseous helium has reasonable cooling capabilities and a large radiation length, important to avoid large multiple scattering, it is the coolant of choice. 
A schematic of the cooling concept indicating the different types of helium flow in the experiment is shown in figure~\ref{fig:cooling_schematic}. 



\begin{figure}[b]
	\centering
	\includegraphics[width = .9\textwidth]{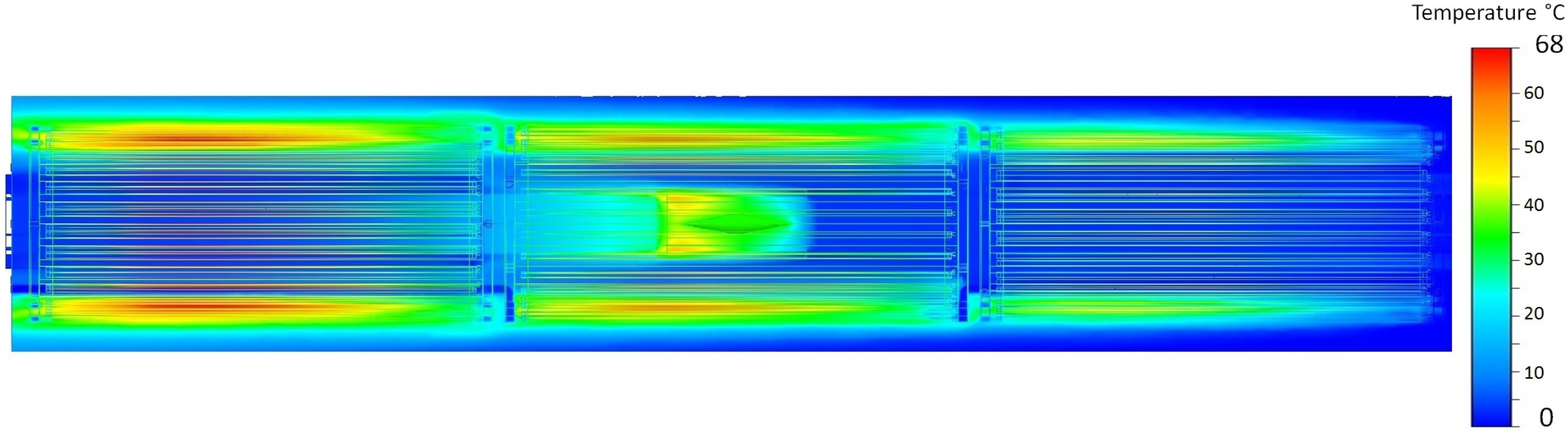}
	\caption{Simulation study of the helium gas cooling with a CAD model of the full detector using the following parameters: power consumption per area $P/A~=~\SI{250}{mW/cm^2}$ and flow velocities $v_{global}~=~\SI{3.5}{m/s}$, $v_{layer1-2}~=~\SI{4}{m/s}$, $v_{layer 3-4}~=~\SI{3.5}{m/s}$ and $v_{fold}~=~\SI{16}{m/s}$.}
	\label{fig:cooling_sim}
\end{figure}

Using a full CAD detector model, we performed computational fluid dynamics studies of the helium cooling~\cite{msc:sam}. 
Figure~\ref{fig:cooling_sim} shows the outcome of a simulation at the targeted sensor power consumption per area of \SI{250}{mW/cm^2}. 
With the flow parameters given in figure~\ref{fig:cooling_sim}, cooling of the detector below \SI{70}{\celsius} seems feasible. 
For higher power consumption, the flow velocities have to be increased. 

\begin{figure}[tb]
	\centering
	\includegraphics[width = .6\textwidth]{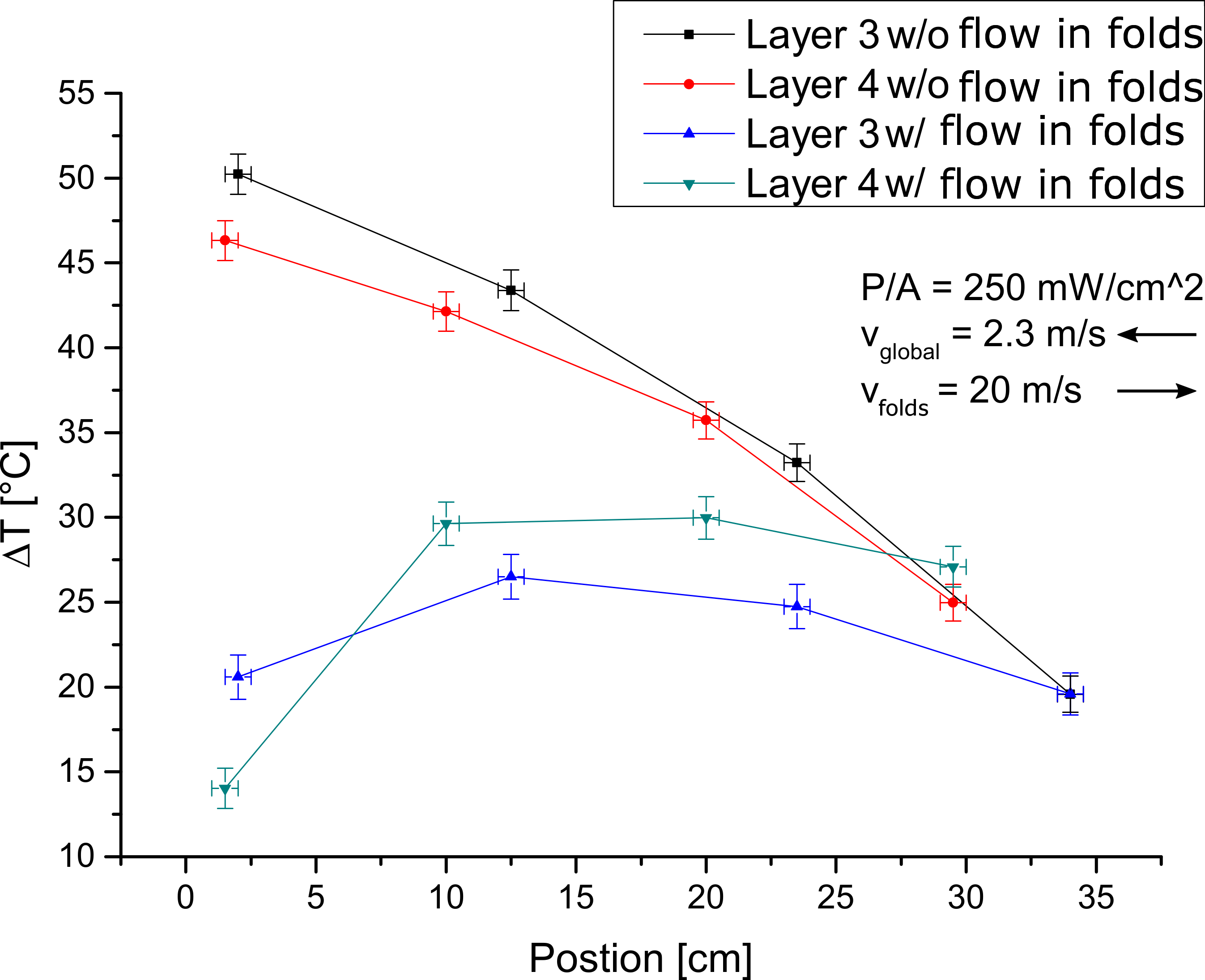}
	\caption{Measured temperature at different positions along a mockup module of layer \num{3} and \num{4}. 
	The temperature is plotted as temperature difference compared to the ambient temperature, around \SI{20}{\celsius}. 
	The mockup was heated with a power per area of \SI{250}{mW/cm^2} and located inside a helium atmosphere with a global flow velocity of $v_{global}~=~\SI{2.3}{m/s}$. 
	The temperature was measured with flow in the folds turned off and on. 
	The flow velocity in the v-shaped folds was $v_{folds}~=~\SI{20}{m/s}$. 
	The arrows represent the flow directions.}
	\label{fig:cooling_meas}
\end{figure}

We performed cooling and mechanical studies in the laboratory with a large scale detector mockup using heatable, in-house manufactured, aluminized-\kapton strips glued to \SI{50}{\micro\metre} thin glass plates. 
Figure~\ref{fig:cooling_meas} shows the measured temperature at different positions along a mockup module while it is heated with a power per area of \SI{250}{mW/cm^2}~\cite{msc:adrian}.
The global helium flow enters to the right of the plot, thus the temperature increases towards position \SI{0}{cm} without helium flowing in the v-shaped folds. 
Using the v-shaped flow channels, the maximum temperature is decreased by more than \SI{20}{\celsius} and the temperature gradient along the module is significantly reduced. 
As the flow velocities, especially for the flow in the folds, are quite high, we tested the mechanical stability of the mockup with a Michelson interferometer. 
We found that it is mechanically very sturdy. 
Vibrations induced by cooling velocities up to \SI{20}{m/s} were found to be below \SI{10}{\micro\metre} which is manageable for the final experiment~\cite{bsc:henkelmann}. 
%

\section{Summary}
\label{sec:sum}

For the Mu3e experiment, it is crucial that the material budget of the tracking detector does not exceed \SI{1}{\permil} of a radiation length per layer. 
In our current design we estimate a peak material budget of about $\SI{1.15}{\permil}\,X_0$ per layer, exceeding this limit slightly.
But in order to achieve proper impedance matching on the FPC, there is not much material left that can be easily reduced. 
We achieved a first technological proof of concept for the FPC using an aluminium FPC prototype by LTU. 
We transferred data at our target data rate of \SI{1.25}{Gb/s} successfully without any bit errors achieving a limit on the bit error rate on the \num{e-13} level. 
The measured conductor thickness of \SI[separate-uncertainty]{12.2(3)}{\micro\metre} allows to improve future FPC designs for the pixel modules. 
Simulation studies and laboratory tests show that a maximum temperature of the Mu3e detector below \SI{70}{\celsius} can be achieved using helium gas cooling. 

\acknowledgments

The authors would like to express their gratitude to the team of LTU for the good cooperation. 
S.~Dittmeier acknowledges support by the \textit{International Max Planck Research School for Precision Tests of Fundamental Symmetries}. 
A.~Herkert acknowledges  support  by  the  HighRR  research  training  group [GRK 2058]. 
N.~Berger acknowledges funding by the \emph{PRISMA cluster of excellence}, Mainz, Germany, and wishes to thank Deutsche Forschungsgemeinschaft for support through the Emmy Noether program.

\end{document}